\documentclass[
floats,
aps,
nofootinbib,
superscriptaddress,
reprint,
prd,
longbibliography]{revtex4-2}

\usepackage{amsmath,amssymb,amsfonts}
\usepackage{mathtools}
\usepackage{hyperref}
\hypersetup{colorlinks,citecolor=blue,urlcolor=blue,linkcolor=blue}
\usepackage{graphicx}
\usepackage{xcolor}
\usepackage{subfigure}
\usepackage{nicefrac}
\usepackage{mathrsfs}
\usepackage{mdframed}
\usepackage{units}
\usepackage{slashed}
\usepackage{wrapfig}
\usepackage{booktabs}

\graphicspath{{plots/}}

\newcommand{\gs}{g_\star}

\newcommand{\Trh}{T_{\rm rh}}
\newcommand{\Tmax}{T_{\rm max}}
\newcommand{\aph}{a_\phi}
\newcommand{\anr}{a_{\rm nr}}
\newcommand{\arh}{a_{\rm rh}}
\newcommand{\acr}{a_{\rm sat}}

\begin{document}
\title{Quantum Seesaw Cosmology: A Heating Phase from Pauli Blocking}

\author{Nicolás Bernal}
\affiliation{New York University Abu Dhabi, PO Box 129188, Saadiyat Island, Abu Dhabi, UAE}
\author{Chee Sheng Fong}
\affiliation{Centro de Ciências Naturais e Humanas, Universidade Federal do ABC, 09.210-170, Santo André, SP, Brazil}

\begin{abstract} 
In \emph{Seesaw Cosmology}, a scalar field reheats the Standard Model (SM) through right-handed neutrinos (RHNs), which generate tiny neutrino masses through the seesaw mechanism. When RHNs rapidly fill a restricted momentum-space region, expansion continually opens new fermionic states that are refilled by subsequent scalar decays, maintaining an approximately constant physical RHN density. This gives rise to a novel heating regime in which the SM temperature increases with the scale factor as $T \propto a^{+3/8}$, in sharp contrast to the conventional $T \propto a^{-3/8}$ scaling during matter-dominated reheating.
\end{abstract}
\maketitle

\section{Introduction}
The standard thermal history of the Universe is observationally tested from Big Bang nucleosynthesis (BBN) onward, whereas its evolution at earlier times remains much less constrained~\cite{Allahverdi:2020bys, Batell:2024dsi}. After cosmic inflation, or more generally after an early period dominated by a scalar field, the energy stored in the dominant component must be transferred to particles that eventually form the Standard Model (SM) thermal bath~\cite{Dolgov:1989us, Traschen:1990sw, Kofman:1994rk, Kofman:1997yn, Amin:2014eta, Barman:2025lvk}. This reheating process establishes the initial conditions for radiation domination and determines the maximum temperature, the expansion history, and the entropy production before BBN. It can therefore have important consequences for the generation of cosmological relics, baryogenesis, dark-matter production, and primordial gravitational-wave backgrounds.

Right-handed neutrinos (RHNs) provide a particularly economical portal between the post-inflationary Universe and the reheating of the SM~\cite{Lazarides:1990huy, Buchmuller:2011mw, You:2024hit, Han:2024qbw, Datta:2025wfh, Mambrini:2026tla, Bernal:2026lpr}. In the type-I seesaw mechanism~\cite{Minkowski:1977sc, Yanagida:1979as, Glashow:1979nm, Gell-Mann:1979vob, Mohapatra:1979ia}, their Majorana masses and Yukawa interactions account for the smallness of the observed active-neutrino masses~\cite{Esteban:2024eli, deSalas:2020pgw}. In \emph{Seesaw Cosmology}, the same particles can mediate the transfer of energy from a scalar field $\phi$ to the SM through the two-step process~\cite{Bernal:2026lpr}
\begin{equation}
    \phi \to N\,N\,, \qquad N \to \ell\, H\,, 
\end{equation}
where $\ell$ and $H$ denote the SM lepton and Higgs doublets, respectively. We refer to $\phi$ as the inflaton for brevity, although the analysis applies more generally to a scalar dominating the pre-radiation era. In the classical dilute treatment, the SM temperature $T$ approaches an approximately constant plateau during scalar domination and subsequently scales with cosmic scale factor $a$ as $T\propto a^{-1/4}$ and $T\propto a^{-3/8}$ during relativistic- and nonrelativistic-RHN domination, respectively~\cite{Bernal:2026lpr}.

However, the classical description implicitly assumes that the RHN occupation numbers remain much smaller than unity. As we will show in this Letter, this approximation can break down even when the microscopic inflaton decay is perturbative. Scalar decays continuously inject RHNs into a restricted region of momentum space, while cosmological expansion redshifts previously produced particles toward lower momenta. If production is sufficiently rapid, the accessible fermionic states can acquire occupations of order one, and Pauli blocking must then be included. At the same time, the inverse reaction $N\, N \to \phi$ can partially return energy to the scalar sector and is Bose enhanced by the occupation of its final scalar state. The relevant quantum statistical factor is therefore
\begin{equation}\label{eq:full_quantum_statitics}
    f_\phi\, (1 - f_{N,1})\, (1 - f_{N,2}) - (1 + f_\phi)\, f_{N,1}\, f_{N,2}\,,  
\end{equation}
where $f_k$ denotes the phase-space distribution of particle $k$. Quantum-statistical effects in nonperturbative fermion production after inflation have been extensively studied in Refs.~\cite{Greene:1998nh, Baacke:1998di, Greene:2000ew, Giudice:1999fb, Peloso:2000hy, Nilles:2001fg}. Finite-density effects, Bose condensation, and Bose-enhanced inflaton decay can also substantially modify perturbative reheating~\cite{Mangano:2001ix, Drewes:2013iaa, Drewes:2014pfa, Bernal:2026dsu}.

In contrast to fermionic-preheating analyzes, we focus on perturbative production from an incoherent population of scalar quasiparticles. This treatment reveals a qualitatively different strong phase-space-filling regime in which the RHN distribution develops a broad, nearly saturated nonthermal momentum band. During scalar domination, expansion continuously moves occupied RHN states away from the injection region, opening new phase-space cells that can be populated by later scalar decays. The RHN number density can consequently remain approximately constant over an extended interval. Since the energy transferred to the SM bath is then sourced at an approximately constant rate per unit volume, while the Hubble expansion rate scales as $H(a)\propto a^{-3/2}$, the sourced SM-radiation component grows as $\rho_R(a)\propto a^{3/2}$, and the SM temperature experiences a novel heating regime following
\begin{equation}
    T \propto a^{+3/8}\,,
\end{equation}
in contrast to the well-known scaling $T\propto a^{-3/8}$ during scalar domination~\cite{Giudice:2000ex}. This rising-temperature phase replaces the approximately constant temperature obtained in the classical dilute regime~\cite{Bernal:2026lpr} whenever strong filling persists throughout scalar domination. Increasing-temperature phases during reheating have previously been found when the dissipation rate of the dominant component increases sufficiently rapidly with the scale factor~\cite{Co:2020xaf}. Here we identify a distinct mechanism arising from the quantum statistics of the particles produced during reheating.

\section{The model} \label{sec:model}
In the seesaw cosmological scenario introduced in Ref.~\cite{Bernal:2026lpr}, after cosmic inflation, the energy density of the Universe is dominated by a real CP-even scalar singlet $\phi$ with mass $m_\phi$. We assume that the scalar $\phi$ has negligible direct couplings to SM fields and transfers its energy through $n$ RHNs $N_i$ with mass $M_i$ of the type-I seesaw model~\cite{Minkowski:1977sc, Yanagida:1979as, Glashow:1979nm, Gell-Mann:1979vob, Mohapatra:1979ia} as follows:
\begin{equation} \label{eq:lag}
    -{\cal L} \supset \frac{m_\phi^2\, \phi^2}{2} + \left[\frac{M_i + \lambda_i\, \phi}{2}\, N_i  \, N_i + y_{i\alpha}\, N_i\, \ell_\alpha\, H + \textrm{H.c.}\right],
\end{equation}
where $H$ and $\ell_\alpha$ $(\alpha=e,\mu,\tau)$ are SM Higgs and lepton $SU(2)_L$ doublets, respectively. We work in the charged-lepton and RHN mass basis and assume, in addition, that the coupling $\lambda$ is diagonal in the RHN mass basis. 

When $\phi$ oscillates around the minimum of its quadratic potential, there are additional time-dependent contributions to the mass of RHNs. We will work in the regime where the mass correction is negligible and in a perturbative regime where nonadiabatic fermion production~\cite{Kofman:1994rk, Shtanov:1994ce, Boyanovsky:1994me, Kofman:1997yn, Greene:1998nh, Greene:2000ew} is not at play, giving respectively,
\begin{equation} \label{eq:small_lambda_Phi}
	|\lambda_i|\, \Phi(t) \ll M_i\,,\qquad
    |\lambda_i|\, \Phi(t) \ll m_\phi\,,
\end{equation}
where $\Phi(t)$ corresponds to the envelope of the oscillating $\phi(t)$. Since there is no explicit $\Phi(t)$ dependence, we will assume that the two bounds are always satisfied by choosing the appropriate $\Phi(t)$. We also assume $m_\phi > 2\, M_i$ for at least one RHN, so that perturbative decay $\phi \to N_i\, N_i$ can take place, with partial decay widths given by
\begin{equation}
	\Gamma_{\phi_i} = \frac{|\lambda_i|^2}{16\pi}\, m_\phi\, \beta_{N_i}^3 \quad \text{where} \quad \beta_{N_i} \equiv \sqrt{1 - \frac{4\, M_i^2}{m_\phi^2}}\,.
\end{equation}
We note that the kinematic condition $m_\phi > 2\, M_i$ makes the second condition of Eq.~\eqref{eq:small_lambda_Phi} a consequence of the first. Cosmic reheating, that is, the formation of the SM thermal bath, is achieved through decays $N_i \to \ell_\alpha\, H$ whose total vacuum decay width can be written as
\begin{equation}
	\Gamma_{N_i} \equiv \frac{m_{\rm eff}\, M_i^2}{8\pi v^2},
\end{equation} 
with $v = 174$ GeV and the effective mass $m_{\rm eff}$ is bounded from below by the solar mass splitting scale $\sqrt{\Delta m^2_{\rm sol}} \simeq \sqrt{7.4 \times 10^{-5}\,{\rm eV}^2}$~\cite{Esteban:2024eli, deSalas:2020pgw} for $n=2$ and normal mass ordering, or by the lightest active-neutrino mass $m_0 \geq 0$ for $n = 3$. 

Due to the fermionic nature of $N_i$, even if their production is perturbative, the RHN occupation numbers can become order one $f_N \sim 1$ and may approach the Pauli limit in part of the distribution when the unblocked production rate exceeds the rate at which expansion opens new phase-space cells.
A schematic estimate is
\begin{equation} \label{eq:saturation_general}
	\Gamma_{\phi_i} \frac{n_\phi}{n_{N_i}^{\rm sat}}\gg {\cal H}\,,
\end{equation}
where ${\cal H}$ is the Hubble expansion rate, $n_\phi$ is the $\phi$ number density and $n_{N_i}^{\rm sat}$ is the maximum number density for $N_i$ allowed by the Pauli exclusion principle.  

In the rest of the work, we follow one RHN species $N \equiv N_i$, assumed to dominate the inflaton decays. We set $M_i \equiv m_N$, $\Gamma_{N_i} \equiv \Gamma_N$, $\beta_{N_i} \equiv \beta_N$, and $\Gamma_{\phi_i} \simeq \Gamma_\phi$. The remaining RHNs generate the rest of the active-neutrino mass spectrum but are either kinematically inaccessible in inflaton decay or not efficiently produced.

\section{Seesaw cosmology with quantum statistics} \label{sec:framework}
Assuming $\phi$ to have a highly nonrelativistic initial distribution given by
\begin{equation} \label{eq:fphi_delta}
    f_\phi(\vec p_\phi, t_0) = (2 \pi)^3\, n_\phi(t_0)\, \delta(\vec p_\phi)\,,
\end{equation}
with $n_\phi$ the number density of $\phi$ and $t_0$ the initial time, the kinetic equations for $\phi$ and $N$ in the Friedmann–Lemaître–Robertson–Walker Universe are, respectively,
\begin{align}
    &\left[\frac{\partial}{\partial t} - {\cal H} p_\phi \frac{\partial}{\partial p_\phi}\right] f_\phi(p_\phi,t) = - \Gamma_\phi f_\phi(p_\phi,t) \left[1 - f_N(p_\star,t)\right] \label{eq:fphib}\\
    &\left[\frac{\partial}{\partial t} - {\cal H}\, p_N\, \frac{\partial}{\partial p_N}\right] f_N(p_N,t) = \frac{2 \pi^2\, \Gamma_\phi\, n_\phi}{p_\star^2}\, \delta(p_N - p_\star) \nonumber \\
    & \qquad\qquad \times \left[1 - f_N(p_\star,t)\right] - \frac{\Gamma_N\, m_N}{E_N}\, f_N(p_N,t), \label{eq:fNb}
\end{align}
with $p_\star \equiv m_\phi\, \beta_N/2$ and where the term $1 - f_N(p_\star,t)$ corresponds to the \emph{Pauli-blocking} factor in the decay of $\phi$. For the RHN decay, the inverse time dilation factor $m_N/E_N$ is crucial and will result in relativistic-RHN domination when $m_\phi \gg m_N$. We have neglected the inverse process $\ell\, H \to N$ as they only have small effects throughout the eras dominated by $\phi$ and $N$. Due to the long-lived nature of $N$, when they start to decay while nonrelativistic, the statistical factor $[1 - f_\ell] [1 + f_H]$ can only result in order one corrections~\cite{Basboll:2006yx, Hahn-Woernle:2009jyb} and therefore has been set to one in our calculations.

Due to the fast SM gauge interactions, the SM can be approximated to be thermalized at each instance, and it is sufficient to track the evolution of its energy density $\rho_R$ through the Boltzmann equation
\begin{equation} \label{eq:rhoR}
    \frac{d\rho_R}{dt} + 4\, {\cal H}\, \rho_R = + \Gamma_N\, m_N\, n_N\,,
\end{equation}
where, due to domination of long-lived $N$, inverse reactions have been neglected. Then ${\cal H}$ is given by the Friedmann equation
\begin{equation} \label{eq:H}
    {\cal H}^2 = \frac{\rho_\phi + \rho_N + \rho_R}{3\, M_P^2}\,,
\end{equation}
where $M_P \simeq 2.4 \times 10^{18}$~GeV is the reduced Planck mass, and $\rho_k$ the energy density of particle $k$ obtained from integrating $E_k f_k$ with $E_k = \sqrt{p_k^2 + m_k^2}$ over its 3-momentum. 

\section{Analytic solutions} \label{sec:anal}
Neglecting the RHN decay and the time evolution in Eq.~\eqref{eq:fNb}, strong saturation occurs for
\begin{equation} \label{eq:Xi}
    \Xi(a) \equiv \frac{f_N(p_\star,a)}{1 - f_N(p_\star,a)} \simeq \frac{2 \pi^2\, \Gamma_\phi\, n_\phi(a)}{{\cal H}(a)\, p_\star^3} \gg 1\,,
\end{equation}
in qualitative agreement with Eq.~\eqref{eq:saturation_general}. During the era of $\phi$ domination, $n_\phi(a) \propto a^{-3}$, ${\cal H}(a) \propto a^{-3/2}$ and then $\Xi(a) \propto a^{-3/2}$ is a monotonically decreasing function. Therefore, strong saturation can occur if $\Xi(a_I) \gg 1$ with $a_I = a(t_0)$ the initial scale factor, which corresponds to
\begin{equation}\label{eq:saturation_condition}
    \frac{\Gamma_\phi}{{\cal H}_I}\,  \gg \frac{\beta_N^3}{48 \pi^2}\frac{m_\phi^4}{M_P^2\, {\cal H}_I^2}\equiv r_\star \,,
\end{equation}
with ${\cal H}_I \equiv {\cal H}(a_I)$ and can be sustained while the condition in Eq.~\eqref{eq:Xi} is satisfied. 

In this fully saturated limit $f_N(p_N,a) \simeq 1$ for $p_\star\, a_I/a\lesssim p_N \lesssim p_\star$ and its number density with $n_N(a_I) =0$ is
\begin{equation}
    n_N(a) \simeq \frac{1}{\pi^2} \int_{p_\star \frac{a_I}{a}}^{p_\star} dp_N\, p_N^2 = n_N^{\rm sat} \left[1 - \left(\frac{a_I}{a}\right)^3\right],
\end{equation}
where $n_N^{\rm sat} \equiv m_\phi^3\, \beta_N^3/(24 \pi^2)$. We note that the RHN number density rapidly approaches a constant saturated value $n_N^{\rm sat}$, which is independent of $\Gamma_\phi$, depending only on the masses of the inflaton and the RHN. The corresponding energy densities are
\begin{align}
\rho_\phi(a) &= m_\phi\, n_\phi(a)\,,\quad
    n_\phi(a) \simeq 3\, \frac{M_P^2\, {\cal H}_I^2}{m_\phi} \left(\frac{a_I}{a}\right)^3\,,\\
    \rho_N(a) &\simeq \frac{1}{\pi^2} \int_{p_\star \frac{a_I}{a}}^{p_\star} dp_N\, p_N^2\, E_N 
    \simeq \frac{m_\phi^4 \beta_N^4}{64 \pi^2} \left[1 - \left(\frac{a_I}{a}\right)^4\right],
\end{align}
where the RHN energy density reaches a plateau that depends only on the masses. 

From Eq.~\eqref{eq:rhoR}, during $\phi$-dominated era, we obtain
\begin{align}
    \rho_R(a) \simeq & \frac{\beta_N^3}{132 \pi^2}\, \frac{\Gamma_N\, m_N\, m_\phi^3}{{\cal H}_I} \left(\frac{a}{a_I}\right)^{3/2} \nonumber\\
    & \times \left[1 - \frac{11}{5} \left(\frac{a_I}{a}\right)^3 + \frac65 \left(\frac{a_I}{a}\right)^{11/2}\right].
\end{align}
The temperature of the SM bath can be extracted from $\rho_R$ taking into account that in thermal equilibrium
\begin{equation} \label{eq:T}
    \rho_R(T) = \frac{\pi^2}{30}\, \gs\, T^4,
\end{equation}
where $\gs(T)$ corresponds to the number of relativistic degrees of freedom that contribute to $\rho_R$. At sufficiently large $a/a_I$, 
\begin{equation}\label{eq:T_38}
    T(a) \simeq \left[\frac{5\, \beta_N^3}{22 \pi^4\, \gs}\, \frac{\Gamma_N\, m_N\, m_\phi^3}{{\cal H}_I}\right]^{1/4} \left(\frac{a}{a_I}\right)^{3/8},
\end{equation}
increasing as $T \propto a^{+3/8}$.

The end of the $\phi$-dominated era is defined as the moment at which $\rho_\phi(\aph) = \rho_N(\aph)$ and corresponds to a scale factor $\aph$ given by
\begin{equation} \label{eq:aphiaI}
    R_3 \equiv \frac{\aph}{a_I} \simeq \left(\frac{192 \pi^2}{\beta_N^4}\, \frac{M_P^2\, {\cal H}_I^2}{m_\phi^4}\right)^{1/3},
\end{equation}
which is independent of $\Gamma_\phi$. 

Considering $\Gamma_\phi/{\cal H}_I \ll r_\star$, one will be in the classical regime where $f_N \ll 1$ with~\cite{Bernal:2026lpr}
\begin{equation}
    R_3^{\rm classical}\simeq \left(\frac52\, \frac{{\cal H}_I}{\Gamma_\phi}\right)^{2/3},
\end{equation}
and is sensitive to $\Gamma_\phi$. In the intermediate regime $\Gamma_\phi/{\cal H}_I \sim r_\star$, the evolution involves a transition from the saturation to the classical regime.

In the saturation regime, the SM bath increases according to the new power law of Eq.~\eqref{eq:T_38} and reaches its maximum temperature $\Tmax$ at $a \simeq a_\phi$ given by
\begin{equation} \label{eq:Tmax}
    \Tmax^4  = T^4(\aph) \simeq \frac{60\, \beta_N}{11 \sqrt{3}\, \pi^3\, \gs}\, M_P\, \Gamma_N\, m_N\, m_\phi\,,
\end{equation}
which is, once again, independent of $\Gamma_\phi$. On the other hand, in the classical regime, the SM temperature will quickly plateau to~\cite{Bernal:2026lpr} 
\begin{equation}\label{eq:Tmax_classical}
    T_{\rm max, classical}^4 \simeq \frac{30}{\pi^2\, \gs}\, \frac{M_P^2\, \Gamma_\phi\, \Gamma_N\, m_N}{m_\phi}\,.
\end{equation}

After the end of the $\phi$-dominated era, the Hubble expansion is dominated by relativistic RHNs that later become nonrelativistic, before decaying into SM particles. The transition from relativistic to nonrelativistic RHNs can be estimated by computing the epoch at which the hardest remaining RHNs satisfy $p_N = m_N$. Taking into account that the highest momentum corresponds to the latest particles produced (that is, at $a = \aph$), the scale factor $\anr$ at the end of the gradual transition is given by
\begin{equation} \label{eq:R2}
    R_2 \equiv \frac{\anr}{\aph} \simeq \frac{m_\phi}{2\, m_N}\, \beta_N\,,
\end{equation}
which is the same as the result of Ref.~\cite{Bernal:2026lpr}.

The end of the reheating era, defined as the onset of SM domination, can be estimated by the equality of the energy densities of $N$ and SM radiation $\rho_N(\arh) = \rho_R(\arh)$, where
\begin{equation} \label{eq:R1}
    R_1 \equiv \frac{\arh}{\anr} \simeq \left(\frac{25}{36 \pi^2}\, \frac{m_N^4}{M_P^2\, \Gamma_N^2}\right)^{1/3},
\end{equation}
and the corresponding reheating temperature 
$\Trh \equiv T(\arh)$
\begin{equation} \label{eq:Trh}
    \Trh^2 \simeq \frac{6}{\pi} \sqrt{\frac{2}{5\, \gs}}\, M_P\, \Gamma_N\,.
\end{equation}
We note that, contrary to the standard case in which the SM bath is created directly from the 
$\phi$, here $\Trh$ depends on $\Gamma_N$ and not on $\Gamma_\phi$~\cite{Buchmuller:2011mw, Mambrini:2026tla, Bernal:2026lpr}.

In the classical regime $\Gamma_\phi/{\cal H}_I \ll r_\star$, while $\Trh$ remains the same, we have~\cite{Bernal:2026lpr}
\begin{equation} \label{eq:R1_summary}
    R_1^{\rm classical}  \simeq \left(\frac{16}{\beta_N^4}\, \frac{m_N^4\, \Gamma_\phi^2}{m_\phi^4\, \Gamma_N^2}\right)^{1/3},
\end{equation}
which depends on $\Gamma_\phi$. Since $R_1 \propto 1/R_3$ and $R_3^{\rm classical} < R_3$, $R_1^{\rm classical}$ is always larger than $R_1$ in the saturation regime. 

\section{Numerical solutions} \label{sec:numerical}
While we have presented the simplified kinetic equations in the previous section, we will now verify our analytic results by solving the momentum-dependent kinetic equations including the full quantum statistical factor of Eq.~\eqref{eq:full_quantum_statitics} for $\phi \leftrightarrow N N$. The detailed numerical implementation is described in Appendix~\ref{app:numerical_treatment}.

For the numerical solution, we choose $\mathcal{H}_I = m_\phi = 10^{13}~{\rm GeV}$, $m_N = 10^9~{\rm GeV}$, $\Gamma_\phi = 10^{11}$~GeV and $\Gamma_N = 10^{-8}$~GeV. 
For this benchmark, the phase-space-filling parameters defined in Eq.~\eqref{eq:Xi} are
\begin{equation}
    \Xi(a_I) \simeq 3 \times10^{11}, \quad \Xi(a_\phi) \simeq 3 \times10^4,
\end{equation}
placing the system deeply in the strong phase-space-filling regime throughout the $\phi$-dominated epoch. 

 \begin{figure}[t!]
     \includegraphics[width=\columnwidth]{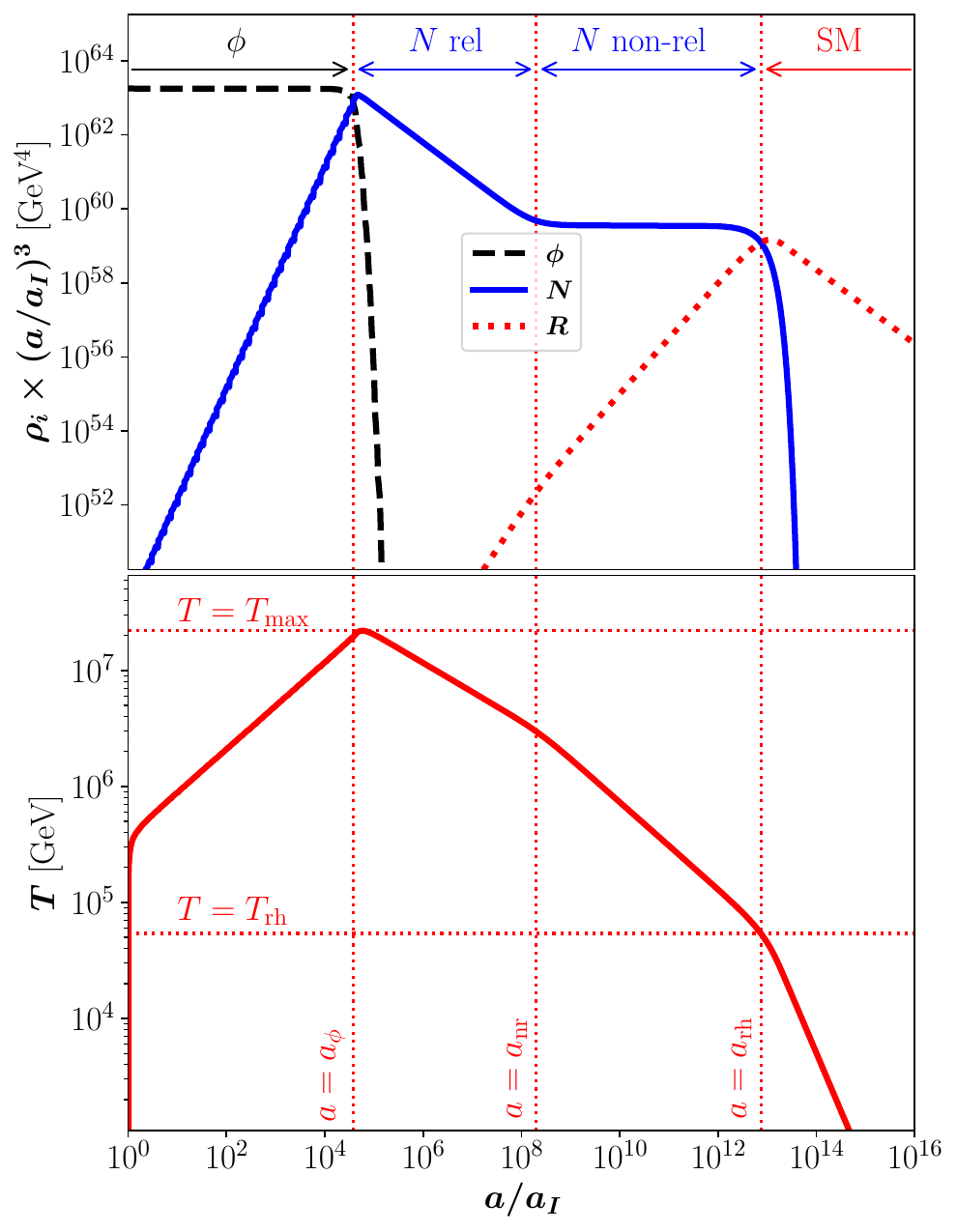}
     \caption{The top plot shows the evolution of the energy densities of $\phi$ (dashed black), $N$ (solid blue) and the SM radiation (dotted red), while the bottom plot shows the evolution of the SM temperature, both as functions of $a/a_I$.}
     \label{fig:evolution}
 \end{figure} 
Figure~\ref{fig:evolution} illustrates the cosmic history for this benchmark. The evolution of energy densities of $\phi$, $N$ and the SM radiation are shown in the top plot. The red dotted vertical lines mark $a = \aph$, $a = \anr$, and $a = \arh$, separating the four asymptotic regimes discussed in the previous section. The bottom plot shows the evolution of the SM temperature, with the horizontal lines indicating $T = \Tmax \simeq 2 \times 10^7$~GeV and $T = \Trh \simeq 5 \times 10^4$~GeV, which fulfills the bound $\Trh > T_{\rm BBN} \simeq 4$~MeV, so as not to spoil the successful predictions of BBN~\cite{Sarkar:1995dd, Kawasaki:2000en, Hannestad:2004px, Barbieri:2025moq}. The numerical transition scales and fitted power laws agree with the analytic estimates from the previous section. 

Notice that the strong-filling condition Eq.~\eqref{eq:saturation_condition} is not sensitive to $\Gamma_N$ as long as $m_\phi \gg m_N$, which will result in relativistic-RHN domination specified by Eq.~\eqref{eq:R2}. In the chosen benchmark which corresponds to $m_{\rm eff} \simeq 8 \times 10^{-12}$~eV, realizable in a three-RHN seesaw model, we have a period of RHN-matter domination, as can be seen in Fig.~\ref{fig:evolution}, in agreement with Eq.~\eqref{eq:R1}. Considering a larger $m_{\rm eff} = \sqrt{\Delta m^2_{\rm sol}}$ consistent with a two-RHN seesaw model with normal mass ordering, RHNs decay promptly when nonrelativistic and the Universe transits from relativistic-RHN domination to SM-radiation domination. Therefore, the cosmic equation of state post scalar domination provides valuable information on the seesaw model.

 \begin{figure}[t!]
     \includegraphics[width=\columnwidth]{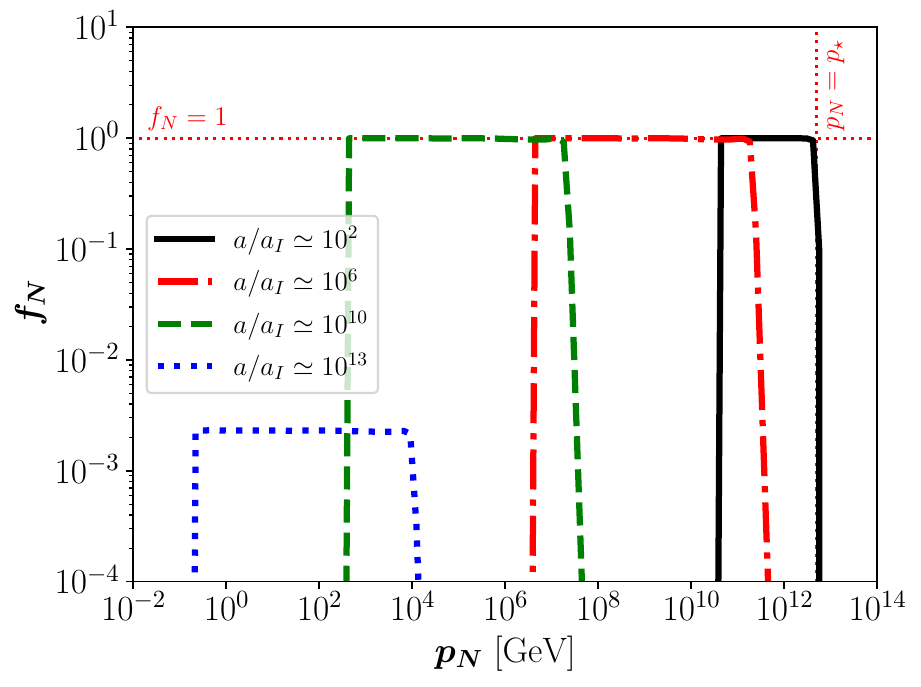}
     \caption{Snapshots of RHN phase-space distribution $f_N(p_N,a)$ at several values of  $a/a_I$, as function of the RHN momentum $p_N$.}
     \label{fig:N_phase_space}
 \end{figure} 
Finally, Fig.~\ref{fig:N_phase_space} shows snapshots of the RHN phase-space distribution $f_N(p_N,a)$ at different stages of cosmological evolution for the benchmark parameters of Fig.~\ref{fig:evolution}. The horizontal red dotted line indicates the Pauli limit, $f_N = 1$, while the vertical red dotted line marks the characteristic injection momentum $p_\star$ for an inflaton at rest. At early times, inflaton decays inject RHNs with momenta close to $p_\star$. Previously produced particles are continuously redshifted toward lower momenta, thereby opening new phase-space cells near the injection scale that can be refilled by subsequent inflaton decays. As a result, the distribution rapidly develops a broad, nearly saturated region with $f_N\simeq1$.

A direct comparison between the evolutions in the strongly Pauli saturated and the formal classical dilute regimes as well as transition from strong filling to the dilute regime, are presented in Appendix~\ref{app:comparison}.

\section{Conclusions} \label{sec:concl}
In this Letter, we have identified a quantum-statistical heating regime during scalar domination in the seesaw cosmology. When fermionic RHNs saturate a restricted region of phase space, expansion continuously opens new momentum states that are rapidly refilled, maintaining an approximately constant physical RHN density. This produces the unusual scaling $T\propto a^{+3/8}$, reaching the maximum temperature $T_{\rm max} \propto (M_P\, \Gamma_N\, m_N\, m_\phi)^{1/4}$ which is independent of $\Gamma_\phi$. The scaling is obtained analytically and confirmed by the full momentum-dependent numerical evaluation. The mechanism applies more generally to two-step reheating through a sufficiently long-lived fermion produced in a restricted momentum region, provided its phase space remains strongly filled and additional interactions do not erase the nonthermal distribution.

The novel heating era will be relevant for phenomena sensitive to the pre-radiation-dominated thermal history. In particular, the abundance and momentum distribution of nonthermally produced RHNs enter the source of nonthermal leptogenesis~\cite{Lazarides:1990huy, Giudice:1999fb, Asaka:1999yd, Asaka:1999jb, Hahn-Woernle:2008tsk, Buchmuller:2011mw, Zhang:2023oyo, Zhang:2025jfh, Datta:2025wfh} through their number density, time-dilated decay rates, and decay epoch. The same altered temperature history may also affect ultraviolet freeze-in, other cosmological relics~\cite{Buchmuller:2011mw, Haque:2023zhb, Cosme:2024ndc, Haque:2024zdq, Datta:2025wfh, Coy:2024itg, Borboruah:2025hai, Mambrini:2026tla, Bernal:2026lpr}, and gravitational-wave signatures associated with the reheating era~\cite{Ringwald:2020ist, Ghoshal:2022kqp, Barman:2023ymn, Barman:2023rpg, Bernal:2023wus, Barman:2024htg, Bernal:2025lxp, Datta:2025wfh, Cline:2026jra}. Finally, the increasing temperature could result in the restoration of a broken symmetry~\cite{Kirzhnits:1972iw, Kirzhnits:1972ut,Dolan:1973qd, Weinberg:1974hy, Kirzhnits:1976ts} as the Universe expands. 

\acknowledgments
NB received funding from the grants PID2023-151418NB-I00 funded by MCIU/AEI/10.13039/ 501100011033/ FEDER and PID2022-139841NB-I00 of MICIU/AEI/10.13039/501100011033 and FEDER, UE. CSF acknowledges the support by Fundação de Amparo à Pesquisa do Estado de São Paulo (FAPESP) Contract No. 2019/11197-6  and Conselho Nacional de Desenvolvimento Científico e Tecnológico (CNPq) under Contract No. 304917/2023-0.

\bibliographystyle{apsrev4-1} 
\bibliography{references.bib}
\onecolumngrid
\appendix
\section{Numerical treatment} \label{app:numerical_treatment}
For a spatially homogeneous and isotropic distribution, we define phase-space distributions $f_\phi(p,t)$ and $f_N(p,t)$ for inflaton and RHN, respectively.\footnote{We define $f_N$ per helicity state and assume equal populations of the two RHN helicities; the matrix-element sums over the final helicities are included.} The unintegrated Boltzmann equation for $f_\phi(p_\phi,t)$ is
\begin{align} \label{eq:fphi}
    \left[\frac{\partial}{\partial t} - {\cal H}\, p_\phi\, \frac{\partial}{\partial p_\phi}\right] f_\phi(p_\phi,t) &= - \frac{1}{2\, E_\phi}\, \frac{1}{2!} \int d\Pi_1\, d\Pi_2\, (2 \pi)^4\, \delta^{(4)}(p_\phi - p_1 - p_2) \left|\mathcal{M}_{\phi \to NN}\right|^2 \nonumber\\
    &\hspace{0.3cm} \times \Big[f_\phi\, [1 - f_N(p_1)]\, [1 - f_N(p_2)] - [1 + f_\phi]\, f_N(p_1)\, f_N(p_2)\Big],
\end{align}
where ${\cal H}$ is the Hubble expansion rate and $E_k = \sqrt{p_k^2 + m_k^2}$ is the energy of the particle species $k$ with mass $m_k$ and momentum $p_k$. The 1/2! is the identical-particle symmetry factor for the two Majorana RHNs. In addition, $d\Pi_k \equiv d^3 p_k/[(2\pi)^3\, 2E_k]$ is the Lorentz-invariant phase-space measure, the factor $(2\pi)^4\, \delta^{(4)}(\cdots)$ corresponds to the conservation of energy and momentum, and $|\mathcal{M}_{\phi \to NN}|^2$ is the squared matrix element for the decay of $\phi$. In the second line, the first term describes Pauli-blocked decay $\phi \to N\, N$, while the second describes Bose-enhanced inverse decay $N\, N\to \phi$.

For one RHN species, the unintegrated Boltzmann equation for its phase-space distribution is
\begin{align} \label{eq:fN}
    \left[\frac{\partial}{\partial t} - {\cal H}\, p_1\, \frac{\partial}{\partial p_1}\right] f_N(p_1,t) &= + \frac{1}{2\, E_1} \int d\Pi_\phi\, d\Pi_2\, (2 \pi)^4\, \delta^{(4)}(p_\phi - p_1 - p_2) \left|\mathcal{M}_{\phi \to NN}\right|^2 \nonumber\\
    &\quad \times \Big[f_\phi\, [1 - f_N(p_1)]\, [1 - f_N(p_2)] - [1 + f_\phi]\, f_N(p_1)\, f_N(p_2)\Big] \nonumber\\
    &\quad - \Bigg[\frac{1}{2\, E_1} \sum_\alpha \int d\Pi_\ell\, d\Pi_H\, (2 \pi)^4\, \delta^{(4)}(p_1 - p_\ell - p_H) \left|\mathcal{M}_{N \to \ell_\alpha H}\right|^2 \nonumber\\
    &\quad \times \Big[f_N\, [1 - f_\ell]\, [1 + f_H] - [1 - f_N]\, f_\ell\, f_H\Big] + \text{CP}\Bigg].
\end{align}
The first two lines are the counterparts of Eq.~\eqref{eq:fphi}, without the 1/2!, while the last two lines correspond to the process $N \leftrightarrow l_\alpha H$, summing over all lepton flavors $\alpha$ and over its CP-conjugate channel.

We solve the momentum-dependent Boltzmann equations using the logarithm of the scale factor
\begin{equation}
    x \equiv \ln\!\left(\frac{a}{a_I}\right),
\end{equation}
as a time variable, and set $a_I = 1$. It is convenient to introduce the comoving momentum
\begin{equation}
    q \equiv a\, p\,.
\end{equation}
In terms of $q$, the Liouville operator associated with the cosmological redshift reduces to
\begin{equation}
    \left[\frac{\partial}{\partial t}  - {\cal H}\, p\, \frac{\partial}{\partial p}\right] f_k(p,t) = {\cal H} \left.\frac{\partial f_k(q,x)}{\partial x}\right|_q,
\end{equation}
for particle species $k$. The collisionless redshifting of the distributions is therefore automatically accounted for, and only the collision terms need to be evolved explicitly.

\subsection{Momentum discretization and initial conditions}
We discretize $q$ on a logarithmic grid with bin boundaries $q_{j-1/2}$ and $q_{j+1/2}$, with $j$ the bin number. For each particle species $k$, we define the phase-space weight
\begin{equation}
    {\cal W}_{k,j} \equiv \frac{g_k}{6\pi^2} \left(q_{j+1/2}^3 - q_{j-1/2}^3\right)
\end{equation}
and the comoving number abundance in a momentum cell stored in that bin,
\begin{equation}
    {\cal U}_{k,j} \equiv {\cal W}_{k,j}\, f_{k,j}\,.
\end{equation}
The bin centers are chosen as the volume-weighted momenta
\begin{equation}
    q_j = \frac34\,  \frac{q_{j+1/2}^4 - q_{j-1/2}^4}{q_{j+1/2}^3 - q_{j-1/2}^3}\,.
\end{equation}

The physical number density, energy density, and pressure are then evaluated as
\begin{align}
    n_k(x) &= \frac{1}{a^3} \sum_j {\cal U}_{k,j}\,,\\
    \rho_k(x) &= \frac{1}{a^3} \sum_j E_{k,j}(x)\, {\cal U}_{k,j}\,,\\
    P_k(x) &= \frac{1}{a^3} \sum_j \frac{p_j^2(x)}{3\, E_{k,j}(x)}\, {\cal U}_{k,j}\,,
\end{align}
where
\begin{equation}
    p_j(x) = \frac{q_j}{a} \qquad \text{and} \qquad E_{k,j}^2(x) = m_k^2+\frac{q_j^2}{a^2}\,.
\end{equation}

The SM radiation component is represented by its comoving energy density
\begin{equation}
    {\cal R}(a) \equiv a^4\, \rho_R(a)\,.
\end{equation}
As initial conditions, at the initial time, we take
\begin{equation}
    f_N(q,0) = 0 \qquad \text{and} \qquad{\cal R}(0) = 0\,.
\end{equation}
The inflaton distribution is initialized as a narrow Gaussian in comoving momentum,
\begin{equation} \label{eq:initial_inflaton_distribution}
    f_\phi(q,0) = A_\phi\, \exp\!\left(-\frac{q^2}{2\, \sigma_q^2}\right),
\end{equation}
whose normalization is fixed by
\begin{equation}
    \rho_\phi(a_I) = \sum_j {\cal W}_{\phi,j}\, E_{\phi,j}(a_I)f_{\phi,j}(a_I) = 3\, M_P^2\,{\cal H}_I^2\,.
\end{equation}
The Gaussian in Eq.~\eqref{eq:initial_inflaton_distribution} is imposed only as an initial condition. Since the calculation is performed in comoving momentum, its physical width subsequently redshifts with the expansion. The value of $\sigma_q$ specifies the initial quasiparticle distribution and is reported together with the remaining numerical parameters.

The upper end of the momentum grid is chosen to be larger than the maximum comoving momentum populated during the integration,
\begin{equation}
    q_{\rm inj}(a) \sim a\, p_\star\,,
    \qquad p_\star = \frac{m_\phi}{2}\, \sqrt{1 - \frac{4\ m_N^2}{m_\phi^2}}\,.
\end{equation}

\subsection{Discretized collision operator}
The collision operator for $\phi\leftrightarrow NN$ is decomposed into channels labeled by the parent momentum bin $j$ and by the decay angle $\mu=\cos\theta_\star$ in the inflaton rest frame. The angular integration is performed using a Gauss--Legendre quadrature with $N_\mu$ nodes.

For an inflaton with physical momentum $p_{\phi,j}=q_j/a$, we define
\begin{equation}
    \gamma_\phi = \frac{E_{\phi,j}}{m_\phi}\,, \qquad \beta_\phi = \frac{p_{\phi,j}}{E_{\phi,j}}\qquad \text{and} \qquad E_\star = \frac{m_\phi}{2}\,.
\end{equation}
The daughter energies in the cosmological frame are
\begin{equation}
    E_{1,2} = \gamma_\phi \left(E_\star \pm \beta_\phi\, p_\star\, \mu\right),
\end{equation}
and their comoving momenta are
\begin{equation}
    q_{1,2} = a\, \sqrt{E_{1,2}^2 - m_N^2}\,.
\end{equation}

Each daughter is distributed between the two neighboring momentum bins. The weights are chosen to conserve their energy at fixed $a$. The same interpolation weights are used to evaluate the daughter occupation numbers entering the quantum-statistical factors. This minimizes mismatches between the interpolation and deposition procedures and improves the numerical detailed-balance properties of the collision operator.

For the full quantum-statistical calculation, the net collision factor for one channel is
\begin{equation} \label{eq:numerical_quantum_bracket}
    {\cal B} = f_\phi\, (1 - f_1)\, (1 - f_2) - (1 + f_\phi)\, f_1\, f_2 = f_\phi\, (1 - f_1 - f_2) - f_1\, f_2\,.
\end{equation}
For a channel corresponding to the parent bin $j$ and angular node $k$, the reaction extent $\xi_{jk}$ during one integration substep is obtained implicitly from
\begin{equation} \label{eq:implicit_reaction_extent}
    \xi_{jk} = h_{jk}\, {\cal B}_{jk}(\xi_{jk}),
\end{equation}
where
\begin{equation}
    h_{jk} = \Delta x_{\rm sub}\, w_k\, {\cal W}_{\phi,j}\, \frac{\Gamma_\phi\, m_\phi}{E_{\phi,j}\, {\cal H}}\,.
\end{equation}
Here $w_k$ is the Gauss--Legendre weight normalized to $\sum_k w_k = 1$. The occupation numbers appearing in ${\cal B}_{jk}$ are evaluated after the tentative transfer $\xi_{jk}$, making Eq.~\eqref{eq:implicit_reaction_extent} nonlinear.

We solve this equation with a safeguarded Newton--bisection algorithm. The allowed interval for $\xi_{jk}$ is bounded by the available number of parent inflatons and by the unoccupied RHN phase-space volume in a forward reaction. For an inverse reaction, it is bounded by the number of RHNs available in the two daughter states. This construction enforces
\begin{equation}
    f_\phi \geq 0 \qquad \text{and} \qquad 0 \leq f_N \leq 1
\end{equation}
without relying on an explicit post-update saturation prescription. We emphasize that the bound $f_N \leq 1$ is enforced in the full-quantum and blocking-only calculations. In the formal dilute control calculation, the Pauli-capacity bound is intentionally removed, allowing the extrapolated occupation to exceed unity, and thereby diagnose the breakdown of the dilute approximation.

For each channel, the update conserves the comoving particle-number relation
\begin{equation}
    \Delta N_N^{\rm com} = -2\, \Delta N_\phi^{\rm com},
\end{equation}
and conserves the instantaneous $\phi \leftrightarrow N\,N$ collision energy up to floating-point and momentum-boundary errors.

\subsection{RHN decay and time integration}
The RHN decay into the SM bath is treated using the time-dilated decay rate. For a momentum bin $j$, the optical depth over an interval $\Delta x$ is
\begin{equation}
    \tau_{N,j} = \frac{\Gamma_N\, m_N}{E_{N,j}\, {\cal H}}\, \Delta x\,.
\end{equation}
The comoving RHN population is advanced using the exact exponential solution,
\begin{equation}
    {\cal U}_{N,j}\, (x + \Delta x) = {\cal U}_{N,j}(x)\, e^{-\tau_{N,j}}.
\end{equation}
The energy removed from the RHNs is deposited in the radiation component,
\begin{equation}
    \Delta{\cal R} = a \sum_j E_{N,j} \left[{\cal U}_{N,j}^{\rm old} - {\cal U}_{N,j}^{\rm new}\right].
\end{equation}
This update reproduces
\begin{equation}
    \dot\rho_R + 4\, {\cal H}\, \rho_R = \Gamma_N\, m_N\, n_N
\end{equation}
in the continuum limit.

The evolution is performed with a uniform step in $x$. We use a symmetric split-step arrangement with oppositely ordered half collision sweeps consisting of
\begin{enumerate}
    \item a half-step of RHN decay into radiation,
    \item a half-step collision sweep over the $\phi\leftrightarrow NN$ channels in ascending order,
    \item a half-step collision sweep in reverse order,
    \item a second half-step of RHN decay.
\end{enumerate}

The opposite collision-sweep orderings reduce the dependence on the arbitrary ordering of the discrete reaction channels. The Hubble rate and the particle energies are evaluated at the midpoint of each scale-factor step. At every recorded time, ${\cal H}$ is given by Eq.~\eqref{eq:H} and the SM temperature is obtained from Eq.~\eqref{eq:T}. In the present implementation, $\gs(T) = 106.75$ is taken to be constant. Integration continues until a fixed maximum value of $a/a_I$ is reached or until the radiation fraction exceeds a chosen threshold for a prescribed number of $e$-folds. Distribution snapshots may be stored at arbitrary values of $a/a_I$.

Unless otherwise stated, the numerical runs use
\begin{equation}
    N_q=180\,, \qquad N_\mu=12\,, \qquad \Delta x=0.02\,, \qquad \frac{\sigma_q}{m_\phi}=10^{-2}\,.
\end{equation}
For the final figures, these values are increased or decreased to verify convergence.

The present calculation includes full quantum statistics for $\phi\leftrightarrow NN$, but treats the transfer $N \to \ell\, H$ using the time-dilated decay width. It therefore neglects the inverse reaction $\ell\, H \to N$, the quantum statistics factors of Higgs and the lepton in this process, and thermal corrections to $\Gamma_N$. These effects can be incorporated into a future extension of the numerical framework.

\section{Comparison with the classical dilute approximation} \label{app:comparison}
For $N$, the Pauli-blocking term $1-f_N$ plays a crucial role. Using the same benchmark point as in Fig.~\ref{fig:evolution}, we will show that by dropping $1-f_N$, $f_N$ quickly grows beyond unity, violating the Pauli exclusion principle. As a result, one incorrectly predicts a shorter $\phi$ dominance era and higher SM temperatures before the completion of the reheating. We emphasize that the formal dilute solution is shown only as a diagnostic of the breakdown of the dilute approximation and should not be interpreted as a physically admissible fermionic evolution after $f_N$ reaches order unity.
 \begin{figure}[t!]
     \def\sepf{0.49}
     \includegraphics[width=\sepf\columnwidth]{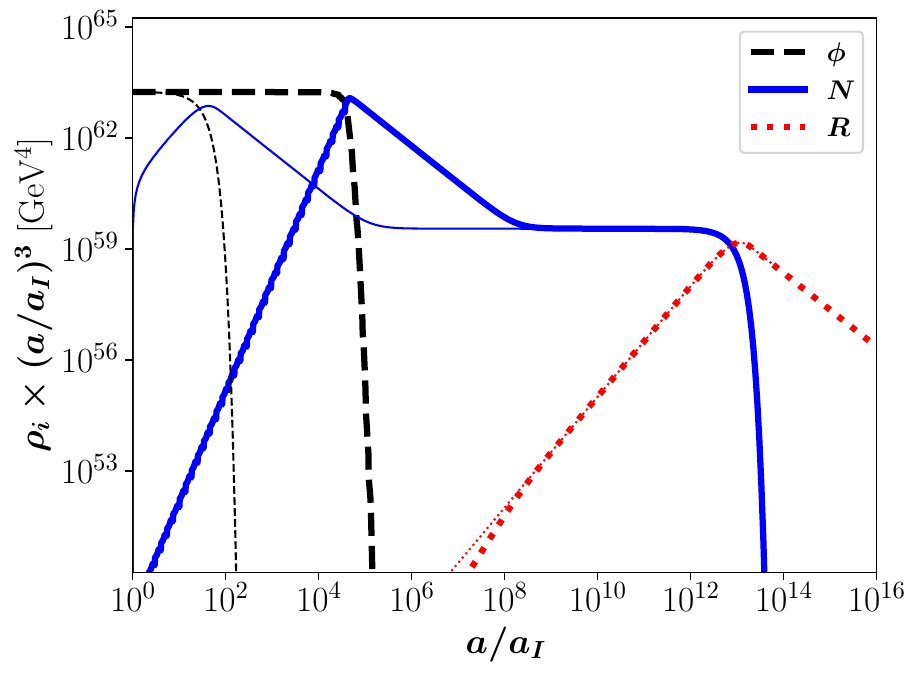}
     \includegraphics[width=\sepf\columnwidth]{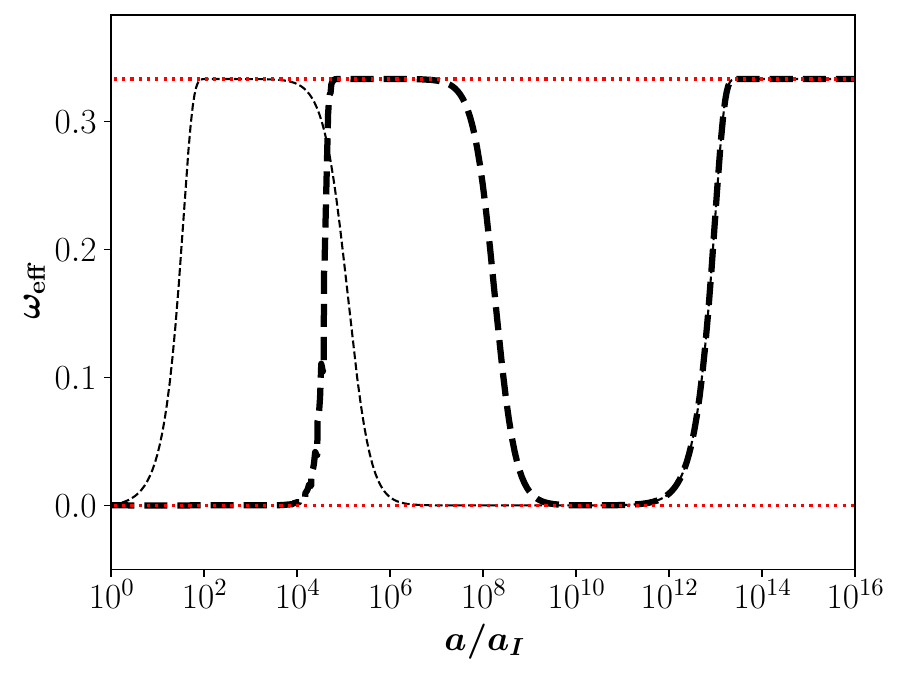}
     \includegraphics[width=\sepf\columnwidth]{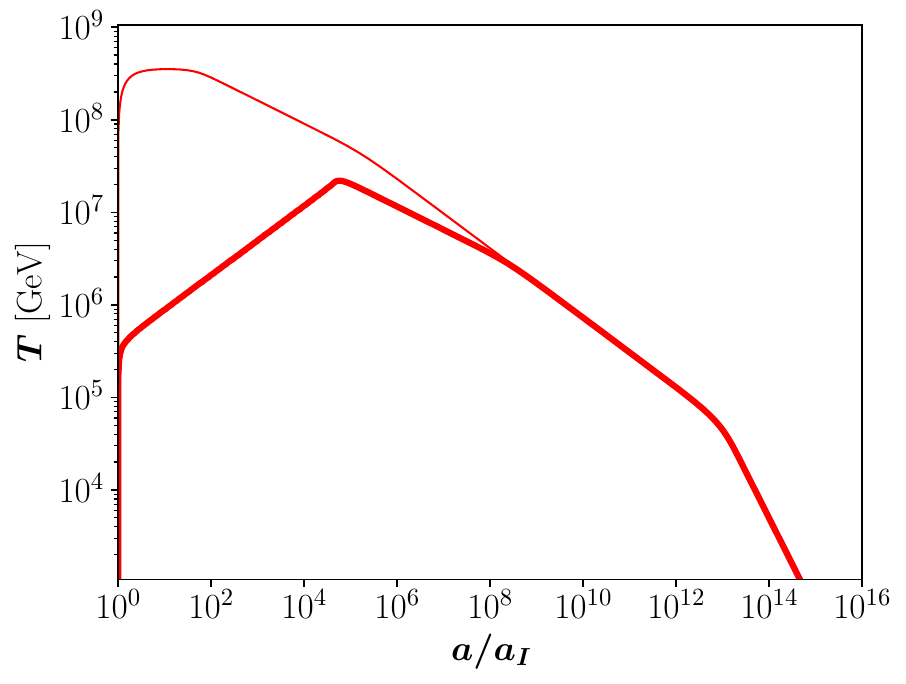}
     \includegraphics[width=\sepf\columnwidth]{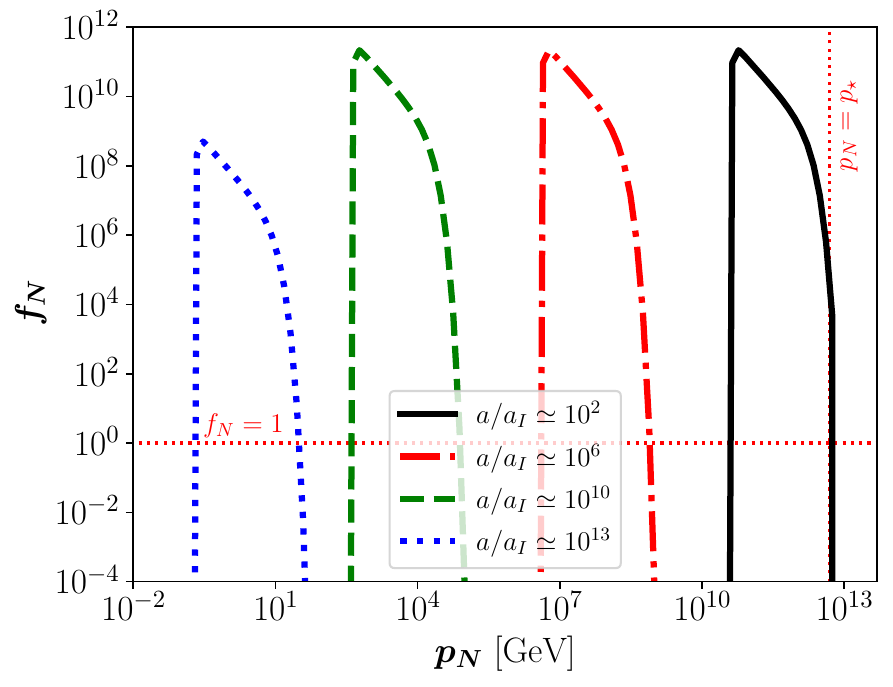}    
     \caption{Comparison of the cosmic evolution obtained with the full quantum-statistical collision term for $\phi \leftrightarrow N\, N$ (thick curves) and in the classical dilute approximation (thin curves), for the benchmark parameters of Fig.~\ref{fig:evolution}. In the lower-right panel, we show the formal RHN occupation obtained in the classical dilute approximation.  In this control calculation, both the Pauli factors $(1 - f_N)\, (1 - f_N)$ and the inverse process $N\, N \to \phi$ are omitted.}
     \label{fig:figAb}
 \end{figure} 
Figure~\ref{fig:figAb} isolates the combined impact of Pauli blocking and inverse production by comparing the full phase-space calculation, shown with thick curves which are identical to that of Fig.~\ref{fig:evolution}, while the thin curves show the result dropping both Pauli blocking and the inverse reaction. In the top-right plot of Fig.~\ref{fig:figAb}, we show, as a black dashed curve, the evolution of the cosmic effective equation of state defined as
\begin{equation}
    \omega_{\rm eff} = \frac{P_\phi + P_N + \rho_R/3}{\rho_\phi + \rho_N + \rho_R}\,,
\end{equation}
where $P_k$ is the pressure obtained by integrating $p_k^2 f_k/(3E_k)$ over its 3-momentum.
For reference, the red horizontal dotted lines indicate $\omega=0$ for matter domination and $\omega=1/3$ for radiation domination. The figure clearly exhibits the smooth sequence of the matter-, radiation-, matter-, and radiation-dominated eras.

Without the Pauli-blocking term, inflatons decay efficiently at early times, rapidly producing a relativistic RHN population with $f_N \gg 1$ as shown in the lower-right panel of Fig.~\ref{fig:figAb} and generating a comparatively hotter SM bath. As a result, the $\phi$ domination era is shorter and the SM temperature quickly plateaus to the maximal value before decreasing again when all $\phi$ have decayed.

At later times, after inflaton production has become negligible, both calculations recover the relativistic- and nonrelativistic-RHN eras, followed by the final SM-radiation-dominated epoch. Their late equality temperatures are similar because they are mainly controlled by the $N$ decay width $\Gamma_N$, although normalizations and transition scale factors retain information about the earlier evolution. The lower-right panel makes the effect particularly transparent: the characteristic sequence of matter-, radiation-, matter-, and radiation-like expansions is present in both cases, but quantum statistics substantially delays the onset of the intermediate RHN-dominated eras.

\subsection*{Transition from strong filling to the dilute regime}
\begin{figure}[t!]
     \def\sepf{0.49}
     \centering
     \includegraphics[width=\sepf\columnwidth]{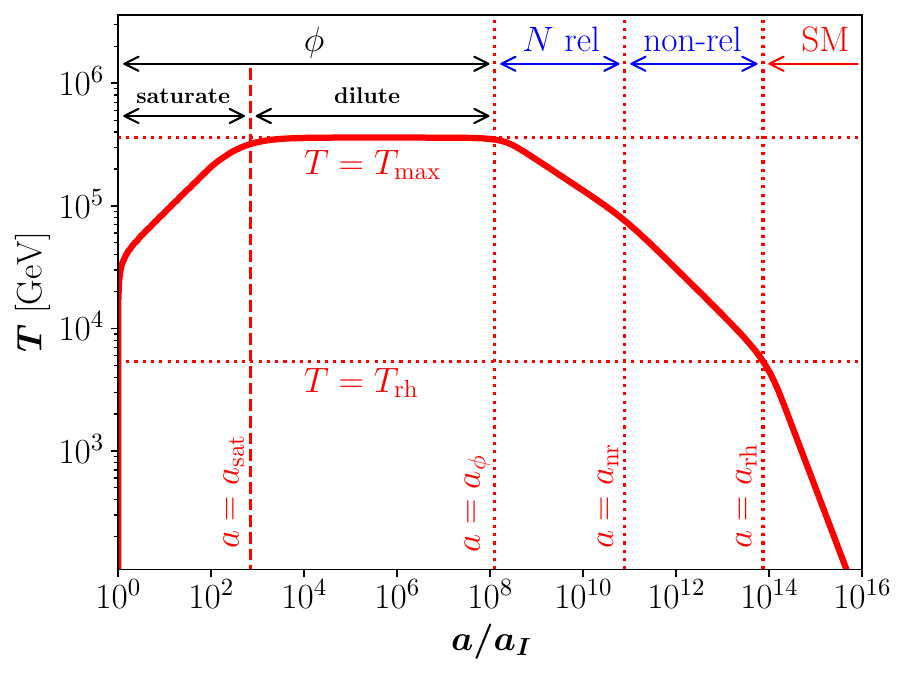}
     \includegraphics[width=\sepf\columnwidth]{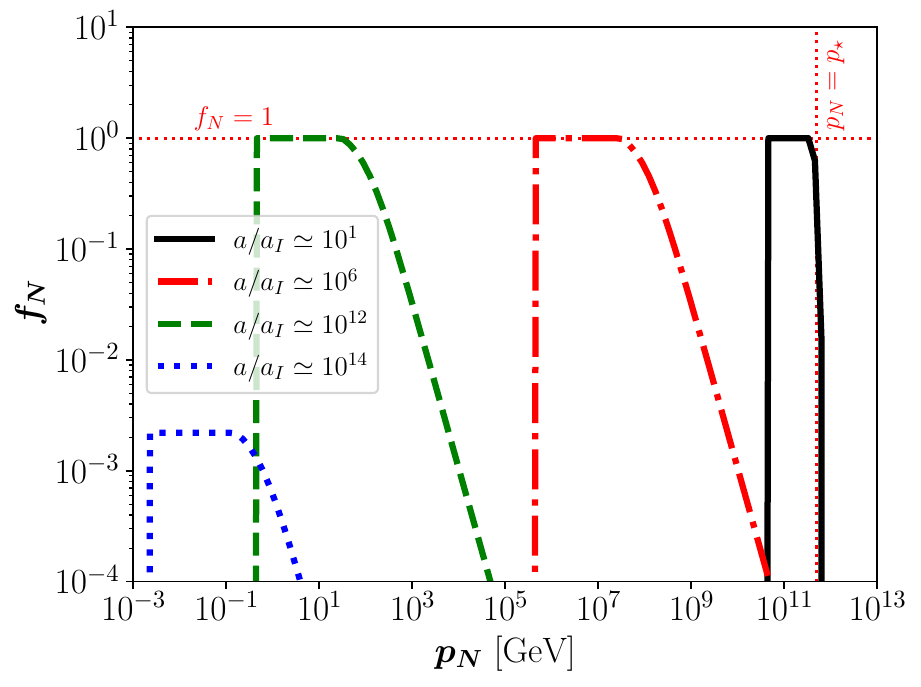}
     \caption{Evolution of the SM temperature (left) and snapshots of the phase-space distribution of the RHN (right) for ${\cal H}_I = m_\phi = 10^{12}$~GeV, $m_N = 10^9$~GeV, $\Gamma_\phi = 1$~GeV and $\Gamma_N = 10^{-10}$~GeV.}
     \label{fig:figD}
 \end{figure}

Finally, we comment on the possibility of having an intermediate regime in which the system crosses over out of the strongly filled regime during the inflaton-dominated period; that is, $a_I < \acr < \aph$. Figure~\ref{fig:figD} shows the evolution of the SM temperature (left) and snapshots of the phase-space distribution of the RHN (right) for ${\cal H}_I = m_\phi = 10^{12}$~GeV, $m_N = 10^9$~GeV, $\Gamma_\phi = 1$~GeV and $\Gamma_N = 10^{-10}$~GeV. This corresponds to the filling parameters $\Xi(a_I) \simeq 3 \times 10^3 \gg 1$ and $\Xi(\aph) \simeq 3 \times 10^{-9} \ll 1$ which reflect the fact that the inflaton-dominated era starts strongly saturated but ends in the dilute regime. After a short initial transient, the temperature approaches the $T(a) \propto a^{+3/8}$ strong-filling solution. Following the crossover near $a = \acr$, it relaxes toward the dilute temperature plateau. Similarly, for a given value of the scale factor, only the low-momentum part of the distribution remains saturated (that is, $p_\star\, a_I/a \lesssim p_N \lesssim p_\star\, \acr/a$), which were produced at early times.

\end{document}